# Unsupervised learning of ferroic variants from atomically resolved STEM images

Mani Valleti,[1] Sergei V. Kalinin,[2] Christopher T. Nelson,[2] Jonathan J. P. Peters,[3] Wen Dong,[3] Richard Beanland,[3] Xiaohang Zhang,[4] Ichiro Takeuchi,[4] and Maxim Ziatdinov[2,5]

[1] Bredesen Center for Interdisciplinary Research, University of Tennessee, Knoxville, TN 37996, USA

[2] Center for Nanophase Materials Sciences, Oak Ridge National Laboratory, Oak Ridge, TN 37831

[3] Department of Physics, University of Warwick, Gibbet Hill Road, Coventry CV4 7AL, UK

[4] Department of Materials Science and Engineering, University of Maryland, College Park, MD 20742

[5] Computational Sciences and Engineering Division, Oak Ridge National Laboratory, Oak Ridge, TN 37831

An approach for the analysis of atomically resolved scanning transmission electron microscopy data with multiple ferroic variants in the presence of imaging non-idealities and chemical variabilities based on a rotationally invariant variational autoencoder (rVAE) is presented. We show that an optimal local descriptor for the analysis is a sub-image centered at specific atomic units, since materials and microscope distortions preclude the use of an ideal lattice as a reference point. The applicability of unsupervised clustering and dimensionality reduction methods is explored and are shown to produce clusters dominated by chemical and microscope effects, with a large number of classes required to establish the presence of rotational variants. Comparatively, the rVAE allows extraction of the angle corresponding to the orientation of ferroic variants explicitly, enabling straightforward identification of the ferroic variants as regions with constant or smoothly changing latent variables and sharp orientational changes. This approach allows further exploration of the chemical variability by separating the rotational degrees of freedom via rVAE and searching for remaining variability in the system. The code used in the manuscript is available at https://github.com/saimani5/ferroelectric_domains_rVAE.

**Introduction**

Ferroic phenomena in complex materials have been a primary focus of condensed matter physics and materials science since the dawn of these disciplines. Ferroelectric materials have found a broad spectrum of applications from sensors and actuators to non-volatile memories.[1-4] Ferromagnets underpin multiple energy generation and information technologies[4] while ferroelastic materials such as shape memory alloys are broadly used as structural materials.[5,6] In ferroelectrics and ferroelastics, emergence of the corresponding order parameters is intimately tied to lattice deformation, resulting in a broad spectrum of mechanical coupling effects and enabling a new generation of strain coupled multiferroic and sensing devices.[7]

Over the last two decades, the attention of the scientific community has shifted to the exploration of ferroic behavior on the nanometer and atomic scales. These advances were enabled by the synergy of improved fabrication techniques such as pulsed laser deposition (PLD) and atomic layer deposition (ALD),[8] and advances in high-resolution imaging and characterization. Techniques such as piezoresponse force microscopy and spectroscopy have enabled visualizing ferroelectric domain structures and probing the polarization dynamics on the nanometer level.[9-12] Scanning transmission electron microscopy (STEM) has allowed visualization of atomic structures of surfaces and interfaces with a high information limit, allowing determination of atomic column positions with picometer level precision.[13] Owing to the strong coupling between ferroic order parameters and mechanical and structural phenomena in ferroelectrics and ferroelastics, the direct visualization of corresponding order parameter fields and their evolution at relevant microstructural elements has been enabled.[14-16]

While the first visualization of ferroelectric domains from atomic-scale transmission electron microscopy (TEM) imaging can be traced to early 2000,[17] this direction gained prominence with further resolution gains from the advent of spherical aberration correctors and the demonstration by Jia et al. of the direct mapping of polarization fields at interfaces and domain walls by TEM.[15,18,19] Almost immediately, a similar approach was introduced using STEM by Chisholm,[16] Borisevich,[20-22] Nelson and Pan,[23-26] and others, leading to rapid growth in this field over the last decade. The behavior of polarization, octahedral tilts in the image plane and in the beam direction, and chemical and physical strain fields can now be visualized.[27-29] Correspondingly, multiple studies of the polarization behavior at surfaces, interfaces, domain walls, and more complex topological defects such as closure domains and ferroelectric vortices, have been reported.[26] In several cases, the numerical values of polarization fields have been fitted to mesoscopic Ginzburg-Landau type models to extract the gradient and flexoelectric terms in order parameter expansions.[30,31]

However, these analyses have been limited to nearly ideal materials systems imaged under optimal conditions, *e.g.*, single interfaces or domain walls. While multiple observations of polarization fields in the vicinity of complex domain wall structures, impurities, and dislocations have been reported and are routinely available, analyses have been limited due to the large number of possible structural variants affected both by the polarization field distribution, chemical composition, and STEM imaging conditions including sample tilt and beam shape. This problem is common in STEM image analysis and has stimulated the development of machine learning (ML) methods for structural analysis (local crystallography) based on the analysis of atomic neighborhoods,[32-34] sliding window transforms,[35, 36] and the analysis of centered sub-images.[37, 38] These methods differ in the choice of the underpinning physical models used. For instance, sliding window transforms implicitly rely on continuous translational symmetry and use fast fourier transform (FFT) amplitude (rejecting the FFT phase) to extract local periodicities. Local crystallographic analysis necessitates unambiguous atom finding and position refinement, reducing the (S)TEM image to a set of atomic positions. Alternatively, the analysis can be based on sub-images centered on selected atomic features, thereby enabling physics-based descriptors as explored for ferroelectrics in Ref. [[37]] and for the chemically heterogeneous Si-graphene system in Ref. [[38]]

However, the characteristic aspect of ferroic materials is the presence of multiple variants of the same deformed unit cell related though discrete rotational symmetries, which can potentially be affected by the imaging conditions used. This severely limits the applicability of image analysis methods based on convolutional neural networks (CNNs), traditional clustering or dimensionality reduction methods, since each rotational variant will be associated with an individual class in the clustering methods. In the presence of other symmetry lowering distortions or composition variations, these classes are impossible to separate since the presence of additional sources of variability results in interclass mixing and leads to physically undefined structural variants. Here, we demonstrate the applicability of rotationally invariant variational autoencoders (rVAE) for the analysis of ferroic materials in the presence of structural and chemical variants and under non-ideal imaging conditions. We show that the ferroic variant corresponding to different orientations can be readily identified and that rVAEs can disentangle the chemical and physical variability. We further show that special care should be taken to identify the presence of instrumental biases, such as those induced by sample tilt variations, necessitating careful tracing of these effects during image acquisition and analysis.

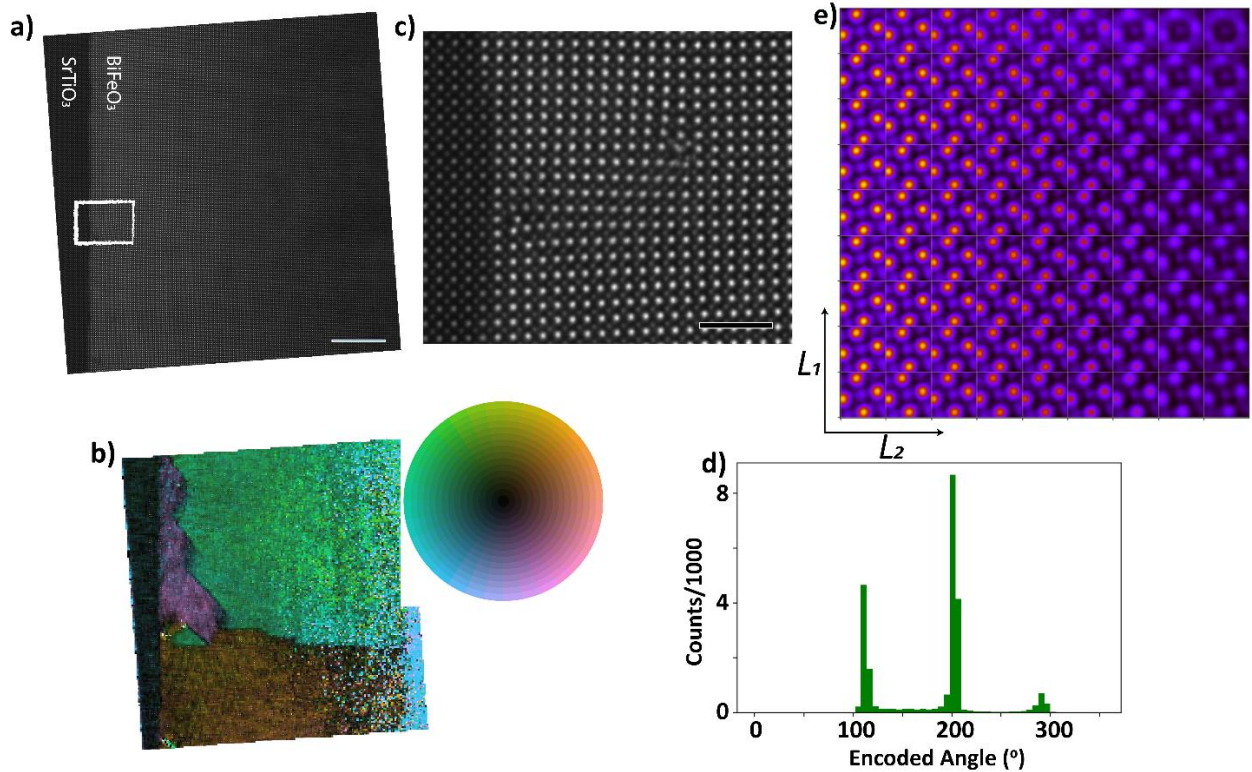

**Figure 1.** (a) HAADF-STEM image of BiFeO$_3$ film. Scalebar 10nm. (b) Corresponding colorized image of cation non-centrosymmetry vector, colorlegend inset. (c) Dissociated <101> edge dislocation into <100> and <001> components enlarged from region highlighted in (a). (d) Histogram of the latent angle distribution. (e) Array of unit cells spanning 2D $L_1$, $L_2$ latent space for the range -2 to 2.

**Results and Discussions**

As a model system we demonstrate rVAE applied to a high-angle annualr dark field (HAADF)-STEM image dataset for a rhombohedral ferroelectric BiFeO$_3$ (BFO) thin film (Figure 1a) grown via PLD. The BFO film adopts a polydomain configuration of the ferroelectric order parameter corresponding to nominal rotationally symmetric twins of the rhombohedral structure. In BFO this order parameter manifests in local cation sublattice as non-centrosymmetry between the 4 corner A-site (Bi) cations and the central B-site (Fe) cation of the perovskite unit cell. This non-centrosymmetry vector at each cation is the offset between the cation position and the average of the four nearest neighbors. The cation positions are plotted as scatter points in real space and are colored with the vector properties of non-centrosymmetry vector using an amplitude-phase color scheme in Fig.1b. The magnitudes of the non-centrosymmetry vectors are normalized to fall in [0,1] and then a color is picked for each vector corresponding to its position in polar space. This color variation in the polar space is shown in the inset of Fig.1b, where the radius of the circle is 1. In addition, the film interface with the SrTiO$_3$ substrate is host to a <101> edge dislocation, which dissociates into separate <100> and <001> components as shown in Figure 1c, enlarged from the highlighted region in Figure 1a.

To explore the order parameter behavior in the BFO system, we extend a previously reported approach based on ML analysis of the structural building blocks.[37, 38] In this method, atoms in the selected sub-lattice are chosen as local reference points and the sub-images centered on these atoms are chosen as descriptors. The sub-images are centered at the experimentally determined atomic coordinates rather than at atoms in an ideally periodic lattice. In this manner, the intrinsic strains and distortions present in the material are relevant only on the length scale of the individual sub-images. Comparatively, using the ideal lattice sites as reference points leads to systematic deviation of the atomic positions from the center of sub-images since in this case the relevant length scale is that of the full image.

Previously, we demonstrated that simple linear unmixing techniques such as principal component analysis (PCA) and non-negative matrix factorization (NMF) applied to thus derived sub-images can be used to explore ferroic distortions in the system, with the PCA eigenvectors providing the dominant distortion modes and the loading maps visualized at respective lattice sites yielding the domain structures and visualizing anomalies at domain walls and interfaces.[37] This approach, however, was shown to create an independent PCA component for each domain variant. However, ferroelectric domains correspond physically to identical distortion patterns differing only in the orientation in the image plane. Hence, while linear unmixing methods allow for separating ferroic variants, the smaller variability corresponding to chemical changes in the system or observational conditions in STEM will be overlooked.

Autoencoders (AE) are a class of neural networks that project the original dataset into a low dimensional latent space through a set of dense or convolutional layers. These latent representations of the dataset are subsequently deconvolved into the original dataset. The reconstruction error i.e., difference between the original dataset and the reconstructed dataset is used to train the AE. Variational autoencoder uses a similar approach but learns a continuous latent space representation of the dataset as opposed to the discrete encoding of the dataset by AEs. In their true sense, the VAEs model the generating distribution of the dataset with the latent layer representing a Bayesian prior distribution.[39] They use a stochastic encoder to approximate the true posterior of the generative model which is the decoder.[40] They achieve this by minimizing a Kullback–Leibler (KL) divergence between the true posterior and the approximated distributions in addition to minimizing a reconstruction loss between the original and the encoded-decoded datasets. rVAEs are an adaptation of VAEs to the dataset with rotational symmetry i.e., same datapoint can be present in the dataset at different orientations. rVAEs explicitly enforce one of the latent variables to represent the angle, two optional latent variables allow for the small offsets in the $x$- and $y$- directions, while the remaining latent variables encode data similar to traditional VAEs. In this manner the rVAE latent space has $1+N$ (pure rotation and no translations) or $3+N$ (rotation plus offsets) elements encoding the orientation of the building blocks, optional offsets allowing for sub-pixel displacements of the true atomic positions from the maximal pixel, and latent variables encoding the characteristic distortions. We have previously applied rVAEs to analyze the dynamics of protein nanorod's self-assembly,[41] explore the dynamic evolution of electron bean induced processes in atomic scale structures,[42] investigate the domain switching pathways in ferroelectric materials visualized by piezoresponse force microscopy,[43] create a bottom-up symmetry analysis Bayesian workflow for atomically-resolved data,[44] and study the dynamic domain evolutions in bias induced phase transitions of $BaTiO_3$.[45]

To analyze the subimages of HAADF-STEM image in Fig.1, an rVAE with symmetric encoder-decoder architecture is used. The encoder has 2 convolutional layers, each comprise 128 kernels/filters of size 3*3 with a leaky-ReLU (slope = 0.1) activation function. At the end of these two convolutional layers, a dense layer is used to reduce the dimensionality of the data to the twice the number of latent dimensions. In a symmetric autoencoder, the decoder's architecture is the mirror image of the encoder. The encoder-decoder pair is trained together to minimize a loss function that is the sum of reconstruction error and KL divergence. As can be observed from the variation of non-centrosymmetry vector across the image, the expected factors of variation in the unit cells present in the image include the presence of ferroelectric domains of dissimilar orientation, presence of non-polar phase in the substrate, emergence behavior at interfaces and domain walls, and instrumental factors. The rVAE analysis has been performed for several sets of possible degrees of freedom *viz.*, i) no translation; two latent variables, ii) translations; one latent variable and iii) translations; 2 latent variables. The full set of analysis is provided in the jupyter notebook attached in github repository. However, since the variability in the input dataset is impossible to independently capture using only a few latent dimensions, in this manuscript, we have only discussed the rVAE with 5 latent dimensions *viz.*, an encoded angle, translations/offsets in X ($O_x$) and Y ($O_Y$), two latent variables $L_1$ and $L_2$.

A stack of sub-images centered on the B site cations was formed as the input the rVAE. For generality, we used the data from the image shown in Figure 1a with a sub-image size of 48 pixels corresponding to 7.5Å or ~1.8-unit cells. Each of these subimages is then represented as a 5D vector in the latent space by rVAE. The resulting distribution of the encoded angle is plotted as a histogram Figure 1d depicting three sharp peaks separated by $\pi/2$, indicative of the successful separation of the ferroic variants based on the in-plane rotations which are expected to be $\pi/2$. The latent space of $L_1$-$L_2$ is uniformly sampled in the range of [-2, 2] while the other latent variables are held at constant. This set of uniformly sampled points in the latent space are then decoded back into the space of subimages and are shown in figure. 1e providing insight into traits within the data encoded in the latent variables ($L_1$ and $L_2$). Overall, the uniformity of the latent space suggests that the network discovered the universal representation of the sub-images, i.e., the individual building blocks.

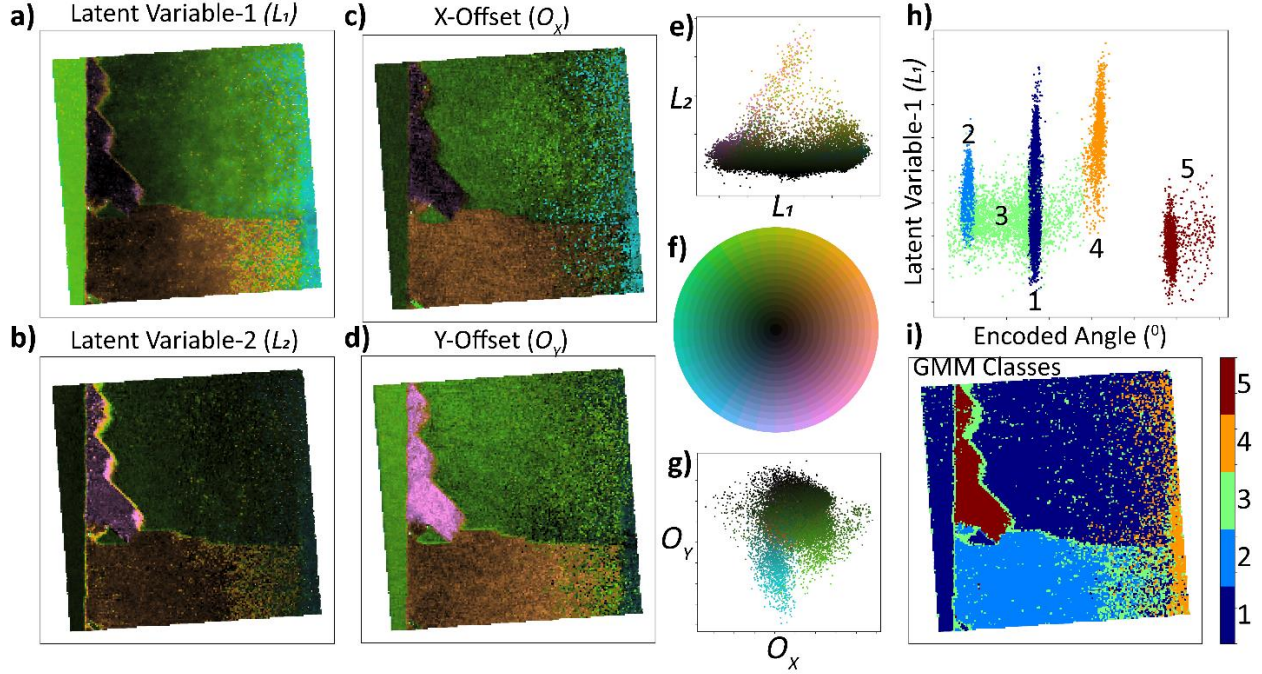

**Figure 2.** Unit cell centers in real space are colored using an amplitude-phase color scheme. The amplitudes/magnitudes of the colors are the (a) Latent variable-1, (b) Latent variable-2, (c) Translations in x-direction, and (d) Translations in y direction encoded by the rVAE, while the angles/phase are the angles encoded by the rVAE in all the four cases. The subimages are represented in the latent space (e) $L_1$ vs. $L_2$ and (g) $O_Y$ vs. $O_x$, and are colored using $L_2$ in (e) and $O_x$ in (g). (f) The colorbar for the amplitude-phase plots where the amplitude varies between [-1, 1]. (h) Unit cells are plotted as scatter points in Latent variable-1 vs Encoded angle space and are colored using the cluster index of GMM. (i) Unit cell centers in the real space are colored using the class index of GMM when the number of classes in GMM is 5.

Greater insight regarding the nature of the rVAE analysis of ferroic phenomena can be derived from exploration of the latent variable maps, as shown in Figure 2. As an initial step to understand the encoding of the subimages, the centers of the B-site cations are plotted as scatter points in the latent space (figure. 2a-d). These points are then colored corresponding to their latent representation using an amplitude phase scheme as discussed for figure. 1b. The corresponding colormap in the polar space is shown in figure. 2f. Amplitudes of the vectors for this representation are the respective encoded values of $L_1$ (figure. 2a), $L_2$ (figure. 2b), $O_X$ (figure. 2c), and $O_Y$ (figure. 2d). The angle/phase for all four plots is the encoded angle of each subimage. Although these latent encodings represented as vectors in the polar form do not represent anything physically, such plots will provide insights about more than one latent dimension in a single plot. It can be observed from figure. 2c-d that the translational encodings along with the angle represent the non-centrosymmetric vector. Two different ferroelectric domains are clearly delineated in both these images and the non-polar substrate is separated in color from the ferroelectric regions. Latent variable-1 ($L_1$) (figure. 2a) along with the angle represents the domain walls to some extent and the chemical variability towards the right end of the sample. Latent variable-2 ($L_2$) extensively represents the domain walls when transitioning from ferroelectric to non-ferroelectric phases.

These unit cell centers are represented in the latent space ($L_2$ vs $L_1$) in figure. 2e and in the latent space ($O_y$ vs $O_x$) in figure. 2g. The colors of these scattered points correspond to the ($L_2$-encoded angle) polar color scheme in figure. 2e and ($O_x$-encoded angle) polar color scheme in figure. 2g.

Insight into the microstructure and evaluation of the latent space segregation can be gained from simple unsupervised classification algorithms, e.g., k-means clustering and Gaussian mixture modeling (GMM). K-means clustering separates the data points based on the selected distance metrics into maximally localized clusters where the distance is the Euclidean distance in the full vector space corresponding to the sub-image size, i.e., the full 2304-dimensional space. GMM represents the probabilistic model for sub-populations (clusters) within an overall population (entire sub-image stack). In GMM, clusters are characterized as mean vectors and covariance matrices in the sub-image space. Unlike distance-based clustering methods, GMM can separate distributions characterized by similar means but with different dispersions. Note that GMM, k-means, and other clustering techniques allow for a significant number of options in terms of distance metrics and form of the covariate matrices. The attached Jupyter notebook allows the reader to experiment with these techniques and apply them to their own data. The 5D latent encodings at each cation site act as the feature vectors (input) to the GMM algorithm. The result of a GGM k=5 cluster analysis is shown in the latent ($L_1$ vs encoded angle) in Figure 2h and real space in (Figure 2i). The scatter points are then colored using the class labels of the clusters formed by GMM. Note, the three ferroelectric domain orientations are assigned to three clusters (#1,#2,#5), one for the poor-fit regions (#4), and one for the domain walls (#3). The class associated with domain walls also encompasses points from the right side of the image that are not part of the domain wall.

To explore the applicability of the rVAE approach for more complex materials systems, we applied the same approach to a $BiFeO_3$−$La_{0.7}Sr_{0.3}MnO_3$ (BFO-LSMO) superlattice epitaxially grown on a $SrTiO_3$ substrate by pulsed laser deposition (PLD).

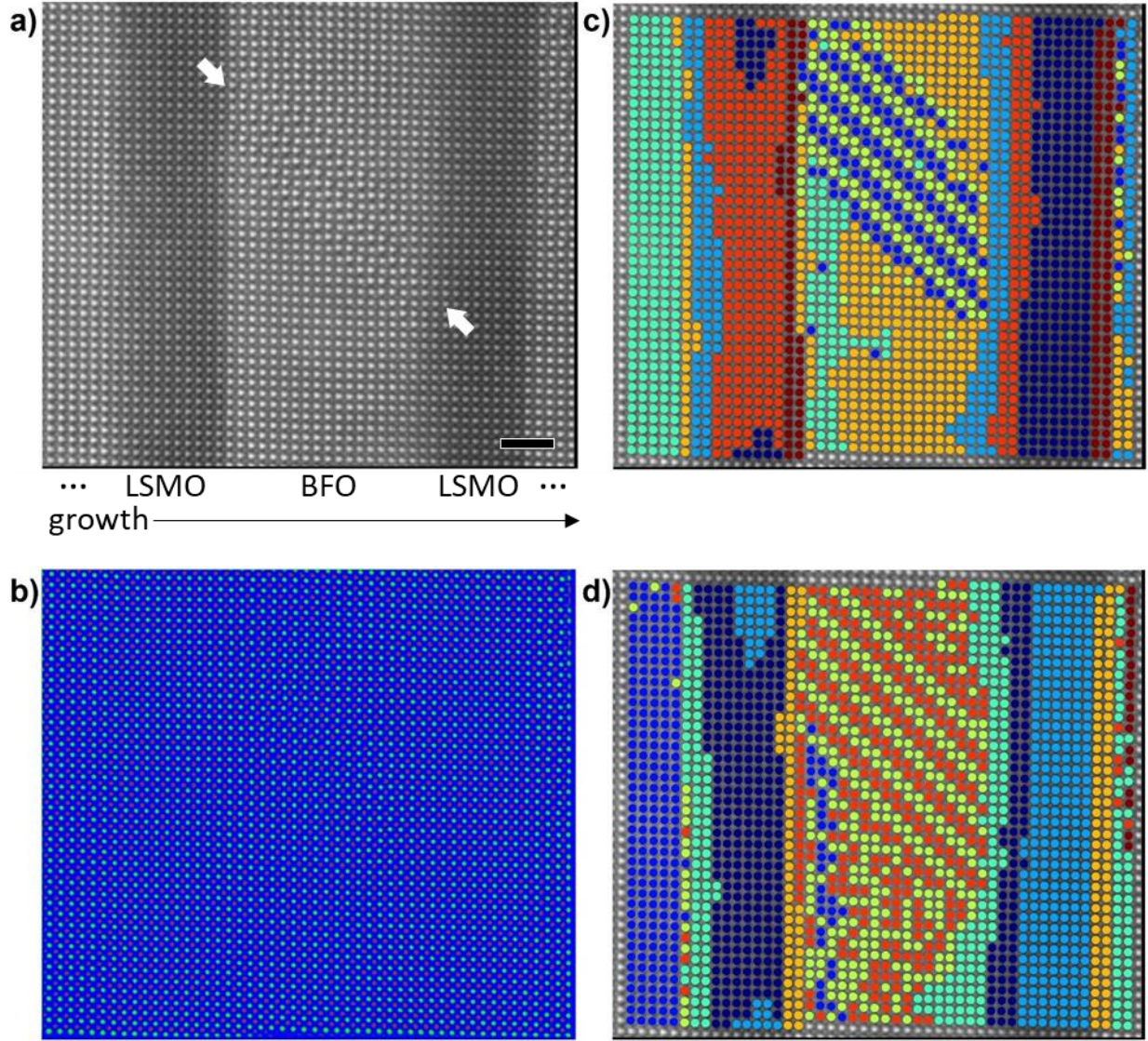

**Figure 3.** (a) ADF-STEM image of BFO-LSMO superlattice. Scalebar 2nm. (b) Output from FerroNET network clearly delineates two different classes of cations (shown by red and green), Cation centers are colored using the labels of their corresponding subimages as a result of (c) GMM and (d) k-means clustering algorithms. Note that the actual value of the class label does not have any significance in this case but the grouping of subimages.

An annular dark field (ADF) STEM image of the BFO-LSMO superlattice acquired along the [100] zone axis is shown in Figure 3a. Here, the bright/dark contrast corresponds to the different BFO/LSMO layers, respectively. Coherent interfaces, i.e., without dislocations, were characteristic in this system; however, the contrast variation is relatively diffuse especially at the BFO-into-LSMO interface, indicative of chemical intermixing at this growth interface. Closer examination of this image (and similar ones) reveals the presence of a defective lamellar structure in some areas within the BFO layers. The lamellae have a periodicity of 2-unit cells and lie 45° with respect to the BFO-LSMO interfaces along (110) direction. Previous image analyses concluded that a highly strained BFO orthorhombic phase coexists with the more common pseudo-rhombohedral phase.[46] This orthorhombic phase presents as the lamella structure with anti-parallel polarization, resulting in an antiferrolectric-like structure.

To illustrate the initial crystallographic analysis of the heterostructures that form within the BFO layers, the A and B cation sites are identified using semantic segmentation via the deep convolutional neural network (DCNN) with the U-Net architecture. The details of this DCNN architecture and its workings are discussed extensively in Ref. [47] The resulting lattice is illustrated in Figure 3b. Note that identification of the atomic sites relies purely on local contrast and no assumptions regarding long-range translational symmetry are made.

As discussed for the previous case, subimages i.e., images cropped around the centers of cations are used as the feature vectors for the cations. Initially, we explored classification of these experimental sub-images into different numbers of classes using two clustering algorithms *viz.*, K-means and GMM. The results of these clustering techniques are shown in figure. 3c for k-means and figure. 3d for the GMM. In all cases, at least eight classes are necessary to visualize domain variants as separate discrete classes, whereas only a few classes separate the variability of chemical composition but not ferroic variants. This is unsurprising since sub-images encode both chemical and polarization effects and the former tend to dominate STEM contrast.

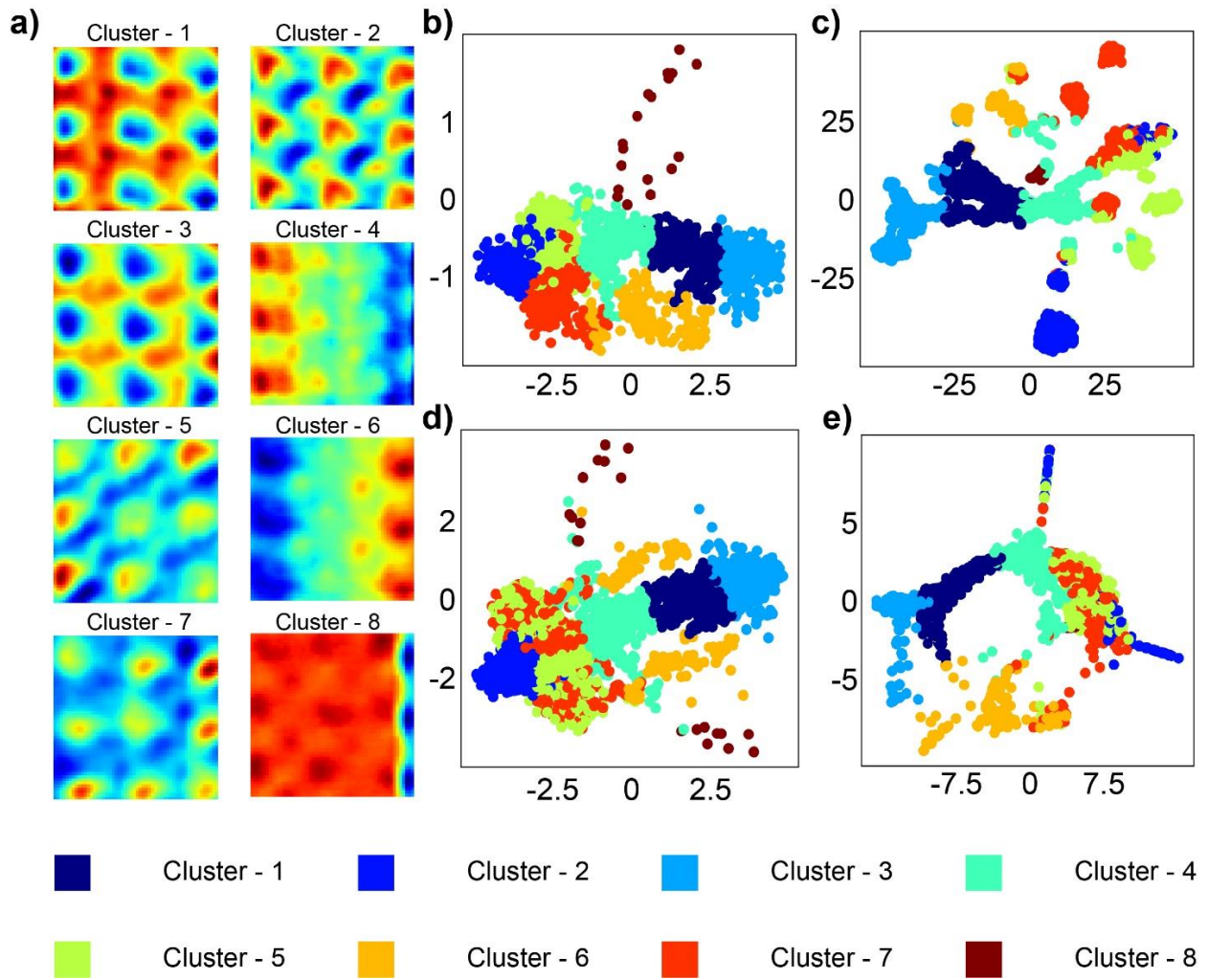

**Figure 4.** (a) Eight cluster centers (means) formed as result of applying GMM on the entire subimage stack. Each sub-image represented as a scatter point in first two components of (b) PCA, (c) tSNE, (d) MDS, and (e) ISO and is color coded using labels of GMM.

This behavior is further illustrated in Figure 4 where the sub-images are visualized in the latent spaces corresponding to different dimensionality reduction methods. Here, we apply classical principal component analysis (PCA), t-distributed stochastic neighbor embedding (tSNE), multidimensional scaling (MDS), and isometric mapping (ISO). This allows us to explore the variability of the first two components, proximity in the data space, distant scaling, and isometric projections for optimal representation of the data manifold in 2D space. These transforms encode the sub-images as component vectors. Unlike clustering methods, this component vector corresponds to the continuous variables. The maps in all cases visualize the distributions of the components in 2D space corresponding to the first and second endmember. Additionally, to build the relationship between the materials structure and dimensionally reduced variables, we color code the GMM classes in the images.

Note that in all cases the distribution of points in the corresponding low dimensional spaces, while structured, are not easily separable. The PCA decomposition forms a dense cloud of the points, and no obvious clusters emerge, the GMM labels suggest considerable intermixing in the PCA space. The density based tSNE decomposition separates clearly visible subclusters, as expected for this method. However, no clear clusters corresponding to the individual components or ferroic variants in the BFO-LSMO system are observed. Similar behaviors are observed for MDS, and ISO based dimensionality reduction. Overall, these methods allow to visualize the clusters within the data but will treat the rotational variants of the same structure as a dissimilar group. Finally, an rVAE with similar architecture discussed for the previous dataset is applied on this dataset of subimage stack with two different window sizes.

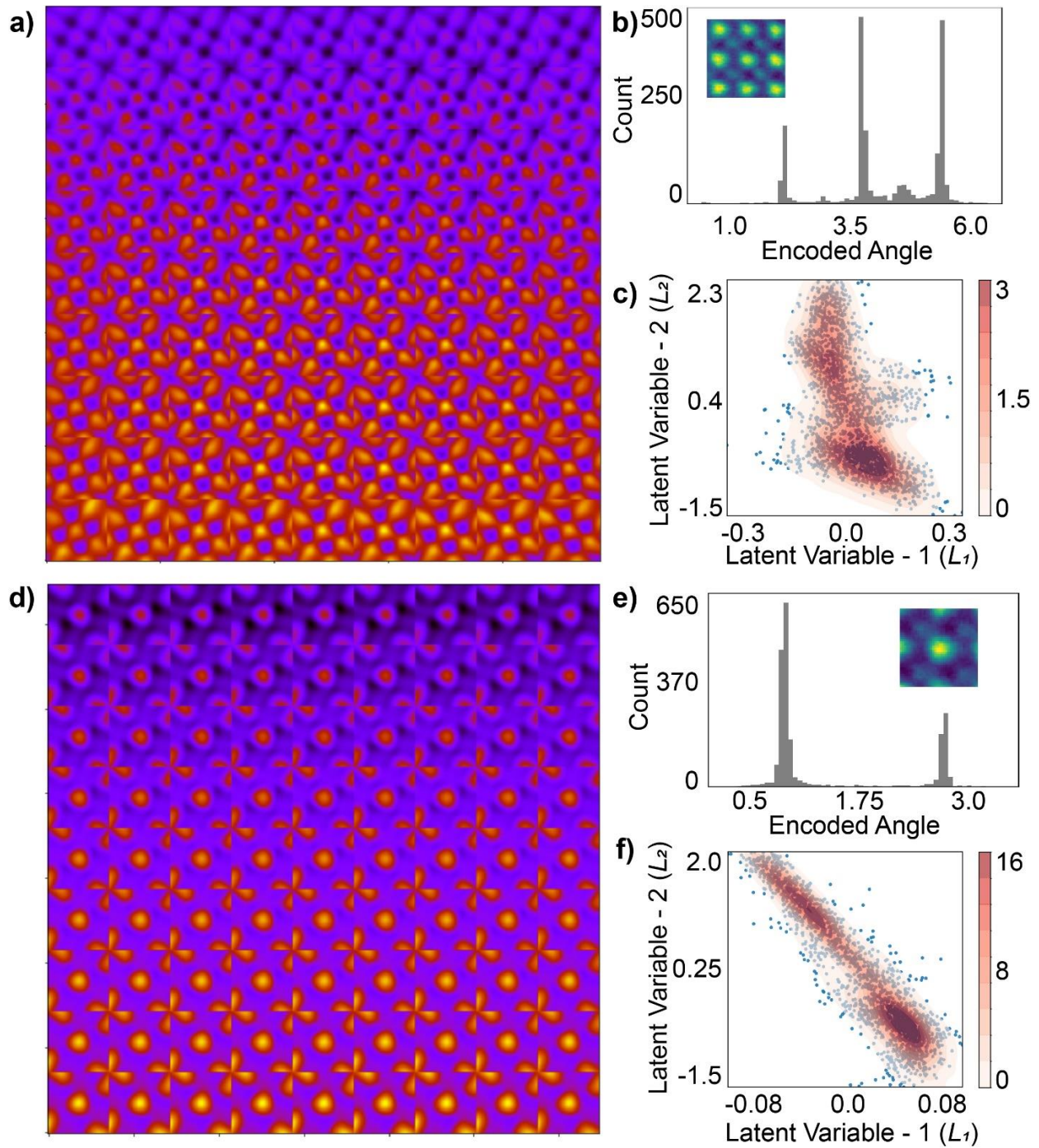

**Figure 5.** Latent space representation (decoding) as a function of first latent variable ($L_1$) and second latent variable ($L_2$) at window sizes (a) 54 pixels and (d) 36 pixels. Histograms of encoded angle at a window size of (b) 54 pixels and (e) 36 pixels. The sample sub-images (feature vectors) for each window size are shown in insets of (b) and (e). Each subimage is plotted as a scatter point in $L_1$-$L_2$ space and the corresponding fitted density of datapoints is shown for window size of (b) 54 pixels and (e) 36 pixels.

The results of the rVAE analysis for two different window sizes are compared in Figure 5. Latent space representation where the space of $L_1$-$L_2$ is uniformly sampled while all other latent variables are held at constants and these sample points are then decoded back to the subimage space is shown in Figure. 5a for window size = 54 pixels and in Figure. 5b for window size = 36 pixels. Similar to the BFO case, the dominant variability in the latent space for window size = 36 pixels is due to the atomic column intensity. This is clearly observed in the joint distribution of data in the $L_1$-$L_2$ space (Figure. 5f) where the $L_2$ varies from -1.5 to 1.5, and $L_1$ varies from -0.08 to 0.05, *i.e.*, 30 times smaller. Furthermore, the latent variable distribution seems to suggest a relationship between the two. The latent angle distribution shows two clear peaks separated by $\pi/2$ as an indication of two ferroelectric domains.

In comparison, for a larger sub-image size the latent space becomes less degenerate, with the range of $L_2$ variability being less than an order of magnitude smaller than that of $L_1$ (Figure. 5c). The structure of the distribution becomes more complex, reflecting more detail in the sub-images. However, examination of the latent space representation suggests that the sub-images contain unphysical features.

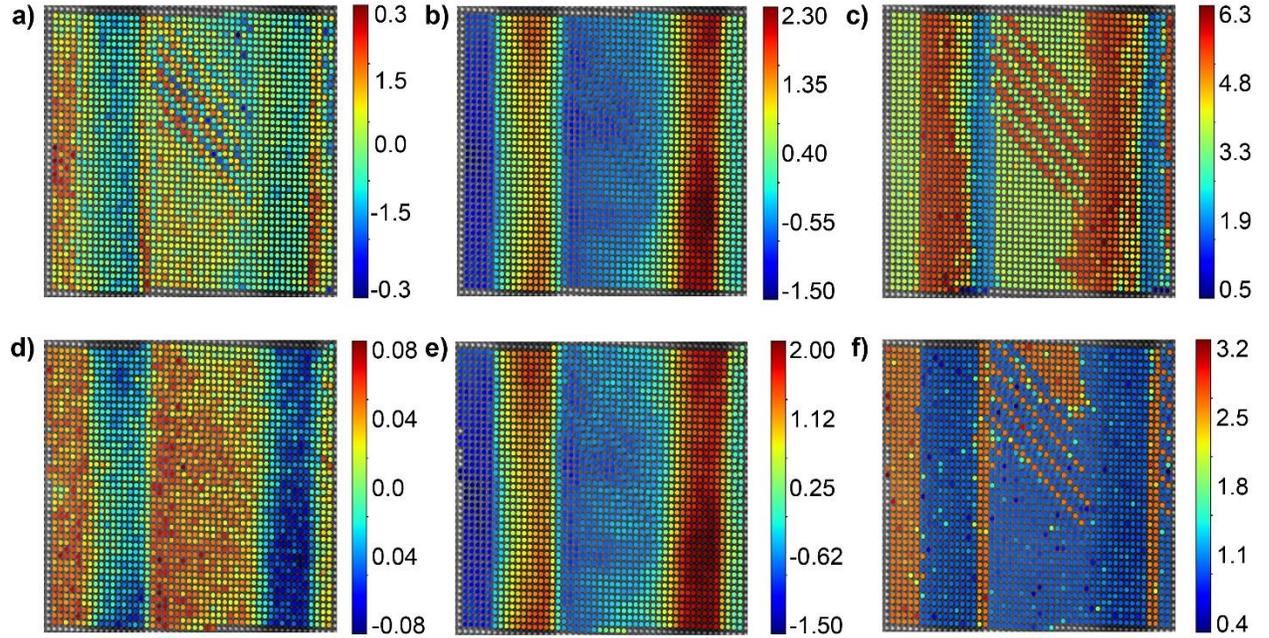

**Figure 6.** Distribution of (a) $L_1$, (b) $L_2$, and (c) encoded angle as a function of position of sub-image for window size of 54 pixels. (d) $L_1$, (e) $L_2$, and (f) encoded angle distributions as a function of spatial position of sub-image at window size of 36 pixels.

The decoded images are shown in Figure 6. Note that for $n = 54$ and $n = 36$ window sizes, the rotation of polarization in the ferroelectric layer, i.e., the antipolar orthorhombic phase, is clearly observed. The $L_2$ images clearly visualize the chemical homogeneity for both window sizes, whereas the corresponding $L_1$ image visualizes similar weak contrast. For the larger window size, $n = 54$, the $L_1$ image now contains information from the ferroelectric structures. In this manner, we have discovered the factors of variability in local atomic structures that visualize ferroic variants and composition variations.

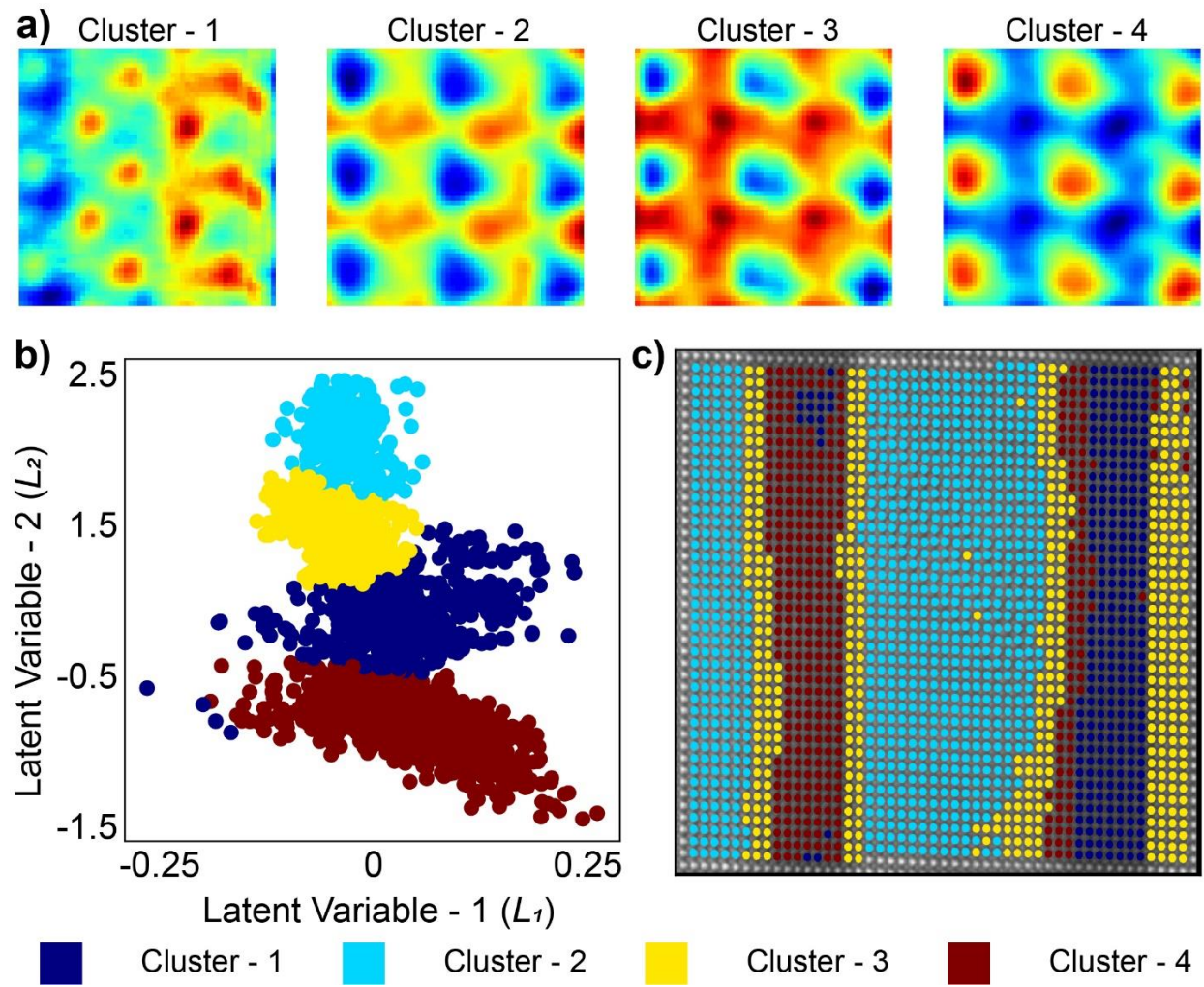

**Figure 7.** (a) Means/Centers of classes as a result of GMM on subimages when represented in encoded latent space ($L_1$ and $L_2$). Each subimage is represented as a scatter point in $L_1$-$L_2$ and color coded with the class index of GMM. (c) Class index of the GMM cluster distribution as a function of spatial positions of the sub-images. Note that orientational variants in cluster 2 are not visible, illustrating good separation between rotational and ferroelectric behaviors.

Note that the analyses in Figures 5 and 6 represent the structural descriptors of the materials in terms of continuous low dimensional variables. While similar to other ML methods, these do not necessarily have well-defined physical meaning; the use of rVAEs allow separation of the rotational variants and remaining degrees of freedom. In the present case, the latter are dominated by chemical composition. This approach further allows exploration of purely chemical behavior of the system. For example, Figure 7 shows GMM clustering on the latent variables. Here, GMM is applied in the latent space of the system where each subimage is represented by its encoding in the 2D latent space with $L_1$, $L_2$ as the latent dimensions. This representation not only helped in the reduction of dimensions from 1296 to 2 but also helps in separating the rotational components of the ferroelectric variants. The resultant cluster centers or means for k = 4 are shown in Figure. 7a. Each subimage is then plotted as a scatter point in $L_1$-$L_2$ latent space and is color coded with its corresponding class label in Figure. 7b. It can be observed from this plot that we can observe four distinct clusters in the latent space, and which is why the number of classes is set to 4 in the first place. Finally, the subimage centers in real space are color coded with their respective class labels in Figure. 7c. Here, the clustering clearly separates the LSMO and BFO regions as well as the intermediate layers between the two, whereas the rotational variants of BFO are identified as a single class.

To summarize, here we explore the applicability of rVAEs for the exploratory analysis of atomically resolved scanning transmission electron microscopy data with multiple ferroic variants in the presence of imaging non-idealities and chemical variability. We show that the optimal local descriptor for the analysis is the sub-image stack centered at specific atomic units, since materials and microscope distortions preclude the use of an ideal lattice being used as the reference point. Classical unsupervised clustering methods such as k-means and GMM tend to produce clusters dominated by chemical and microscope effects and a large number of classes is required to establish the presence of rotational variants. Similarly, dimensionality reduction methods projecting the data on low dimensional spaces via PCA or density-based methods tend to produce non-easily separable distributions.

Comparatively, the rVAE allows extraction of the angle corresponding to the orientation of ferroic variants explicitly, allowing for straightforward identification of ferroic variants as regions with the constant or smoothly changing latent variables and sharp orientational changes. This approach allows for further exploration of chemical variability by separating the rotational degrees of freedom via rVAE and exploring the remaining variability in the system.

This approach can be of interest for systems with more complex polarization structures, including morphotropic ferroelectrics, materials at the morphotropic phase boundaries, and can provide insight into the nature of symmetry breaking phenomena in these materials.

This effort (ML and STEM) is based upon work supported by the U.S. Department of Energy (DOE), Office of Science, Basic Energy Sciences (BES), Materials Sciences and Engineering Division (S.V.K., C.T.N.) and was performed and partially supported (M.Z.) at Oak Ridge National Laboratory's Center for Nanophase Materials Sciences (CNMS), a U.S. Department of Energy, Office of Science User Facility. The work was partly supported by the EPSRC (UK) through grant EP/P031544/1. The support of Marin Alexe and Ana M Sanchez are greatly appreciated. The work at the University of Maryland was supported by ONR MURI N000141310635, ONR MURI N000141712661, and National Institute of Standards and Technology (NIST) Cooperative Agreement 70NANB17H301.

**Materials and methods:**

Sample growth: The oxide layers were grown by pulsed laser deposition (PLD) using a 48 nm wavelength KrF excimer laser. $La_{0.7}Sr_{0.3}MnO_3$ layers were deposited at 600 °C using 0.8 $mJ/cm^2$ with an oxygen pressure of 0.15 mbar and frequency of 5 Hz. $BiFeO_3$ layers were deposited at 600 °C using 0.97 $mJ/cm^2$ with an oxygen pressure of 0.15 mbar and 10 Hz pulse frequency

Atomic-resolution STEM images of the monolithic $BiFeO_3$ film were collected using a probe corrected NION ultraSTEM at 200kV with a high-angle annular dark field (HAADF) detector. The BFO sample preparation, imaging, scan correction, fitting, and image analysis were performed identically to Refs. [48, 49].

Atomic-resolution STEM images of the BFO-LSMO superlattices were acquired using a double aberration-corrected Schottky emission JEOL ARM-200F operating at 200 kV. Annular dark field (ADF) images were formed using a collection angle of 45−180 mrad. To reduce scan distortion and sample drift effect, each image is the result of 20 short exposure images combined using the SmartAlign algorithm [10.1186/s40679-015-0008-4]. TEM specimens were prepared using a FEI Scios focused ion beam (FIB) using standard lift-out procedures.

# References

1. Ha, S. D.; Ramanathan, S., Adaptive oxide electronics: A review. *J. Appl. Phys.* **2011,** *110* (7).
2. Dearaujo, C. A. P.; Cuchiaro, J. D.; McMillan, L. D.; Scott, M. C.; Scott, J. F., FATIGUE-FREE FERROELECTRIC CAPACITORS WITH PLATINUM-ELECTRODES. *Nature* **1995,** *374* (6523), 627-629.
3. Auciello, O.; Scott, J. F.; Ramesh, R., The physics of ferroelectric memories. *Phys. Today* **1998,** *51* (7), 22-27.
4. Waser, R., Nanoelectronics and Information Technology *Nanoelectronics and Information Technology* **2012**.
5. Liu, C.; Qin, H.; Mather, P. T., Review of progress in shape-memory polymers. *J. Mater. Chem.* **2007,** *17* (16), 1543-1558.
6. Salje, E. K. H., Ferroelasticity. *Contemp. Phys.* **2000,** *41* (2), 79-91.
7. A.K. Tagantsev, L. E. C., and J. Fousek, *Domains in Ferroic Crystals and Thin Films*. Springer: New York, 2010.
8. Dawber, M.; Rabe, K. M.; Scott, J. F., Physics of thin-film ferroelectric oxides. *Rev. Mod. Phys.* **2005,** *77* (4), 1083-1130.
9. Kalinin, S. V.; Morozovska, A. N.; Chen, L. Q.; Rodriguez, B. J., Local polarization dynamics in ferroelectric materials. *Reports on Progress in Physics* **2010,** *73* (5), 056502.
10. Gruverman, A. L.; Hatano, J.; Tokumoto, H., Scanning force microscopy studies of domain structure in BaTiO3 single crystals. *Jpn. J. Appl. Phys. Part 1 - Regul. Pap. Short Notes Rev. Pap.* **1997,** *36* (4A), 2207-2211.
11. Soergel, E., Piezoresponse force microscopy (PFM). *J. Phys. D-Appl. Phys.* **2011,** *44* (46), 464003.
12. Kalinin, S. V.; Rar, A.; Jesse, S., A decade of piezoresponse force microscopy: Progress, challenges, and opportunities. *Ieee Transactions on Ultrasonics Ferroelectrics and Frequency Control* **2006,** *53* (12), 2226-2252.
13. Yankovich, A. B.; Berkels, B.; Dahmen, W.; Binev, P.; Sanchez, S. I.; Bradley, S. A.; Li, A.; Szlufarska, I.; Voyles, P. M., Picometre-precision analysis of scanning transmission electron microscopy images of platinum nanocatalysts. *Nat. Commun.* **2014,** *5*.
14. Jia, C. L.; Nagarajan, V.; He, J. Q.; Houben, L.; Zhao, T.; Ramesh, R.; Urban, K.; Waser, R., Unit-cell scale mapping of ferroelectricity and tetragonality in epitaxial ultrathin ferroelectric films. *Nat. Mater.* **2007,** *6* (1), 64-69.
15. Jia, C. L.; Urban, K. W.; Alexe, M.; Hesse, D.; Vrejoiu, I., Direct Observation of Continuous Electric Dipole Rotation in Flux-Closure Domains in Ferroelectric Pb(Zr,Ti)O(3). *Science* **2011,** *331* (6023), 1420-1423.
16. Chisholm, M. F.; Luo, W. D.; Oxley, M. P.; Pantelides, S. T.; Lee, H. N., Atomic-Scale Compensation Phenomena at Polar Interfaces. *Phys. Rev. Lett.* **2010,** *105* (19).
17. Pan, X. Q.; Kaplan, W. D.; Ruhle, M.; Newnham, R. E., Quantitative comparison of transmission electron microscopy techniques for the study of localized ordering on a nanoscale. *J. Am. Ceram. Soc.* **1998,** *81* (3), 597-605.
18. Jia, C. L.; Mi, S. B.; Urban, K.; Vrejoiu, I.; Alexe, M.; Hesse, D., Effect of a Single Dislocation in a Heterostructure Layer on the Local Polarization of a Ferroelectric Layer. *Phys. Rev. Lett.* **2009,** *102* (11).
19. Jia, C. L.; Mi, S. B.; Faley, M.; Poppe, U.; Schubert, J.; Urban, K., Oxygen octahedron reconstruction in the SrTiO(3)/LaAlO(3) heterointerfaces investigated using aberration-corrected ultrahigh-resolution transmission electron microscopy. *Phys. Rev. B* **2009,** *79* (8).
20. Borisevich, A. Y.; Lupini, A. R.; He, J.; Eliseev, E. A.; Morozovska, A. N.; Svechnikov, G. S.; Yu, P.; Chu, Y. H.; Ramesh, R.; Pantelides, S. T.; Kalinin, S. V.; Pennycook, S. J., Interface dipole between two metallic oxides caused by localized oxygen vacancies. *Phys. Rev. B* **2012,** *86* (14).

21. Kim, Y. M.; Morozovska, A.; Eliseev, E.; Oxley, M. P.; Mishra, R.; Selbach, S. M.; Grande, T.; Pantelides, S. T.; Kalinin, S. V.; Borisevich, A. Y., Direct observation of ferroelectric field effect and vacancy-controlled screening at the BiFeO3/LaxSr1-xMnO3 interface. *Nat. Mater.* **2014,** *13* (11), 1019-1025.

22. Borisevich, A. Y.; Eliseev, E. A.; Morozovska, A. N.; Cheng, C. J.; Lin, J. Y.; Chu, Y. H.; Kan, D.; Takeuchi, I.; Nagarajan, V.; Kalinin, S. V., Atomic-scale evolution of modulated phases at the ferroelectric-antiferroelectric morphotropic phase boundary controlled by flexoelectric interaction. *Nat. Commun.* **2012,** *3*.

23. Nelson, C. T.; Winchester, B.; Zhang, Y.; Kim, S. J.; Melville, A.; Adamo, C.; Folkman, C. M.; Baek, S. H.; Eom, C. B.; Schlom, D. G.; Chen, L. Q.; Pan, X. Q., Spontaneous Vortex Nanodomain Arrays at Ferroelectric Heterointerfaces. *Nano Lett.* **2011,** *11* (2), 828-834.

24. Das, S.; Tang, Y. L.; Hong, Z.; Goncalves, M. A. P.; McCarter, M. R.; Klewe, C.; Nguyen, K. X.; Gomez-Ortiz, F.; Shafer, P.; Arenholz, E.; Stoica, V. A.; Hsu, S. L.; Wang, B.; Ophus, C.; Liu, J. F.; Nelson, C. T.; Saremi, S.; Prasad, B.; Mei, A. B.; Schlom, D. G.; Iniguez, J.; Garcia-Fernandez, P.; Muller, D. A.; Chen, L. Q.; Junquera, J.; Martin, L. W.; Ramesh, R., Observation of room-temperature polar skyrmions. *Nature* **2019,** *568* (7752), 368-+.

25. Hong, Z. J.; Damodaran, A. R.; Xue, F.; Hsu, S. L.; Britson, J.; Yadav, A. K.; Nelson, C. T.; Wang, J. J.; Scott, J. F.; Martin, L. W.; Ramesh, R.; Chen, L. Q., Stability of Polar Vortex Lattice in Ferroelectric Superlattices. *Nano Lett.* **2017,** *17* (4), 2246-2252.

26. Yadav, A. K.; Nelson, C. T.; Hsu, S. L.; Hong, Z.; Clarkson, J. D.; Schlepuetz, C. M.; Damodaran, A. R.; Shafer, P.; Arenholz, E.; Dedon, L. R.; Chen, D.; Vishwanath, A.; Minor, A. M.; Chen, L. Q.; Scott, J. F.; Martin, L. W.; Ramesh, R., Observation of polar vortices in oxide superlattices. *Nature* **2016,** *530* (7589), 198-+.

27. Borisevich, A.; Ovchinnikov, O. S.; Chang, H. J.; Oxley, M. P.; Yu, P.; Seidel, J.; Eliseev, E. A.; Morozovska, A. N.; Ramesh, R.; Pennycook, S. J.; Kalinin, S. V., Mapping Octahedral Tilts and Polarization Across a Domain Wall in BiFeO(3) from Z-Contrast Scanning Transmission Electron Microscopy Image Atomic Column Shape Analysis. *Acs Nano* **2010,** *4* (10), 6071-6079.

28. Borisevich, A. Y.; Chang, H. J.; Huijben, M.; Oxley, M. P.; Okamoto, S.; Niranjan, M. K.; Burton, J. D.; Tsymbal, E. Y.; Chu, Y. H.; Yu, P.; Ramesh, R.; Kalinin, S. V.; Pennycook, S. J., Suppression of Octahedral Tilts and Associated Changes in Electronic Properties at Epitaxial Oxide Heterostructure Interfaces. *Phys. Rev. Lett.* **2010,** *105* (8).

29. He, Q.; Ishikawa, R.; Lupini, A. R.; Qiao, L.; Moon, E. J.; Ovchinnikov, O.; May, S. J.; Biegalski, M. D.; Borisevich, A. Y., Towards 3D Mapping of BO6 Octahedron Rotations at Perovskite Heterointerfaces, Unit Cell by Unit Cell. *Acs Nano* **2015,** *9* (8), 8412-8419.

30. Borisevich, A. Y.; Morozovska, A. N.; Kim, Y. M.; Leonard, D.; Oxley, M. P.; Biegalski, M. D.; Eliseev, E. A.; Kalinin, S. V., Exploring Mesoscopic Physics of Vacancy-Ordered Systems through Atomic Scale Observations of Topological Defects. *Phys. Rev. Lett.* **2012,** *109* (6).

31. Li, Q.; Nelson, C. T.; Hsu, S. L.; Damodaran, A. R.; Li, L. L.; Yadav, A. K.; McCarter, M.; Martin, L. W.; Ramesh, R.; Kalinin, S. V., Quantification of flexoelectricity in PbTiO3/SrTiO3 superlattice polar vortices using machine learning and phase-field modeling. *Nat. Commun.* **2017,** *8*.

32. Belianinov, A.; He, Q.; Kravchenko, M.; Jesse, S.; Borisevich, A.; Kalinin, S. V., Identification of phases, symmetries and defects through local crystallography. *Nat. Commun.* **2015,** *6*.

33. Ziatdinov, M.; Banerjee, A.; Maksov, A.; Berlijn, T.; Zhou, W.; Cao, H. B.; Yan, J. Q.; Bridges, C. A.; Mandrus, D. G.; Nagler, S. E.; Baddorf, A. P.; Kalinin, S. V., Atomic-scale observation of structural and electronic orders in the layered compound alpha-RuCl3. *Nat. Commun.* **2016,** *7*.

34. Lin, W. Z.; Li, Q.; Belianinov, A.; Sales, B. C.; Sefat, A.; Gai, Z.; Baddorf, A. P.; Pan, M. H.; Jesse, S.; Kalinin, S. V., Local crystallography analysis for atomically resolved scanning tunneling microscopy images. *Nanotechnology* **2013,** *24* (41).

35. Vasudevan, R. K.; Ziatdinov, M.; Jesse, S.; Kalinin, S. V., Phases and Interfaces from Real Space Atomically Resolved Data: Physics-Based Deep Data Image Analysis. *Nano Lett.* **2016,** *16* (9), 5574-5581.
36. Vasudevan, R. K.; Belianinov, A.; Gianfrancesco, A. G.; Baddorf, A. P.; Tselev, A.; Kalinin, S. V.; Jesse, S., Big data in reciprocal space: Sliding fast Fourier transforms for determining periodicity. *Appl. Phys. Lett.* **2015,** *106* (9).
37. Ziatdinov, M.; Nelson, C.; Vasudevan, R. K.; Chen, D. Y.; Kalinin, S. V., Building ferroelectric from the bottom up: The machine learning analysis of the atomic-scale ferroelectric distortions. *Appl. Phys. Lett.* **2019,** *115* (5), 5.
38. Ziatdinov, M.; Dyck, O.; Jesse, S.; Kalinin, S. V., Atomic Mechanisms for the Si Atom Dynamics in Graphene: Chemical Transformations at the Edge and in the Bulk. *Adv. Funct. Mater.* **2019,** *29* (52), 8.
39. Kingma, D. P.; Welling, M., An Introduction to Variational Autoencoders. *Found. Trends Mach. Learn.* **2019,** *12* (4), 307.
40. Kingma, D. P.; Welling, M., Auto-Encoding Variational Bayes. *arXiv:1312.6114* **2013**.
41. Kalinin, S. V.; Zhang, S.; Valleti, M.; Pyles, H.; Baker, D.; De Yoreo, J. J.; Ziatdinov, M., Disentangling Rotational Dynamics and Ordering Transitions in a System of Self-Organizing Protein Nanorods via Rotationally Invariant Latent Representations. *ACS Nano* **2021,** *15* (4), 6471-6480.
42. Kalinin, S. V.; Dyck, O.; Jesse, S.; Ziatdinov, M., Exploring order parameters and dynamic processes in disordered systems via variational autoencoders. **2021,** *7* (17), eabd5084.
43. Kalinin, S. V.; Steffes, J. J.; Liu, Y.; Huey, B. D.; Ziatdinov, M., Disentangling ferroelectric domain wall geometries and pathways in dynamic piezoresponse force microscopy via unsupervised machine learning. *Nanotechnology* **2021,** *33* (5), 055707.
44. Kalinin, S. V.; Oxley, M. P.; Valleti, M.; Zhang, J.; Hermann, R. P.; Zheng, H.; Zhang, W.; Eres, G.; Vasudevan, R. K.; Ziatdinov, M., Deep Bayesian local crystallography. *npj Computational Materials* **2021,** *7* (1), 181.
45. Ignatans, R.; Ziatdinov, M.; Vasudevan, R.; Valleti, M.; Tileli, V.; Kalinin, S. V., Latent Mechanisms of Polarization Switching from In Situ Electron Microscopy Observations. *Advanced Functional Materials* **2022,** *n/a* (n/a), 2100271.
46. Dong, W.; Peters, J. J. P.; Rusu, D.; Staniforth, M.; Brunier, A. E.; Lloyd-Hughes, J.; Sanchez, A. M.; Alexe, M., Emergent Antipolar Phase in BiFeO3-La0.7Sr0.3MnO3 Superlattice. *Nano Lett.* **2020,** *20* (8), 6045-6050.
47. Ziatdinov, M.; Dyck, O.; Maksov, A.; Li, X. F.; San, X. H.; Xiao, K.; Unocic, R. R.; Vasudevan, R.; Jesse, S.; Kalinin, S. V., Deep Learning of Atomically Resolved Scanning Transmission Electron Microscopy Images: Chemical Identification and Tracking Local Transformations. *Acs Nano* **2017,** *11* (12), 12742-12752.
48. Ziatdinov, M.; Nelson, C. T.; Zhang, X.; Vasudevan, R. K.; Eliseev, E.; Morozovska, A. N.; Takeuchi, I.; Kalinin, S. V., Causal analysis of competing atomistic mechanisms in ferroelectric materials from high-resolution scanning transmission electron microscopy data. *npj Computational Materials* **2020,** *6* (1), 127.
49. Nelson, C. T.; Vasudevan, R. K.; Zhang, X.; Ziatdinov, M.; Eliseev, E. A.; Takeuchi, I.; Morozovska, A. N.; Kalinin, S. V., Exploring physics of ferroelectric domain walls via Bayesian analysis of atomically resolved STEM data. *Nat. Commun.* **2020,** *11* (1), 6361.